\documentclass[a4paper,superscriptaddress,aps,prb,twocolumn,floatfix,citeautoscript,showpacs]{revtex4}

\usepackage{mathtools}  
\mathtoolsset{showonlyrefs,showmanualtags}  
\usepackage{graphicx}
\usepackage{color}
\usepackage{amsmath}
\usepackage{amssymb}
\usepackage[english]{babel}
\usepackage[utf8]{inputenc}

\newcommand{\tobe}[1]{{\color{black} #1}}
\usepackage{bm}
\usepackage{url}
\newcommand{\I}{\mathrm{i}}
\usepackage[colorlinks=true,linkcolor=blue,citecolor=blue,urlcolor=blue]{hyperref}
\newcommand{\ket}[1]{\left| #1 \right>} 
\newcommand{\bra}[1]{\left< #1 \right|} 
\graphicspath{{figures/}}
\newcommand{\vect}[1]{\mathbf{#1}}
\makeatletter
\g@addto@macro\bfseries{\boldmath}
\makeatother

\begin{document}
%
%
\title{Photoemission Spectra from Reduced Density Matrices: the Band Gap in Strongly Correlated Systems}

\author{Stefano Di Sabatino}
\email[]{disabatino@irsamc.ups-tlse.fr}
\affiliation{Laboratoire de Physique Th\'eorique, CNRS, IRSAMC, Universit\'e Toulouse III - Paul Sabatier, 118 Route de Narbonne, 31062 Toulouse Cedex, France and European Theoretical Spectroscopy Facility}
\author{J.A. Berger}
\affiliation{Laboratoire de Chimie et Physique Quantiques, IRSAMC, Universit\'e Toulouse III - Paul Sabatier, CNRS, 118 Route de Narbonne, 31062 Toulouse Cedex, France and European Theoretical Spectroscopy Facility}
\author{Lucia Reining}
\affiliation{Laboratoire des Solides Irradi\'es, \'Ecole Polytechnique, CNRS,
CEA, Université Paris-Saclay, F-91128 Palaiseau, France and European Theoretical Spectroscopy Facility}
\author{Pina Romaniello}
\affiliation{Laboratoire de Physique Th\'eorique, CNRS, IRSAMC, Universit\'e Toulouse III - Paul Sabatier, 118 Route de Narbonne, 31062 Toulouse Cedex, France and European Theoretical Spectroscopy Facility}

\begin{abstract}
We present a method for the calculation of photoemission spectra in terms of reduced density matrices. 
We start from the spectral representation of the one-body Green's function $G$, whose imaginary part is related to photoemission spectra, and we introduce a frequency-dependent effective energy that accounts for all the poles of $G$. Simple approximations to this effective energy give accurate spectra in model systems in the weak as well as strong correlation regime. 
In real systems reduced density matrices can be obtained from reduced density-matrix functional theory. Here we use this approach to calculate the photoemission spectrum of bulk NiO: our method yields a qualitatively correct picture both in the antiferromagnetic and paramagnetic phases, contrary to mean-field methods, in which the paramagnet is a metal.
\end{abstract}
\date{\today}
\maketitle
\section{Introduction}
Photoemission is a powerful tool to obtain insight into the electronic structure and excitations in materials. 
The interpretation of the experimental data is, however, a complicated task. Theory represents, hence, an essential tool for the analysis of the experiments as well as prediction of material properties. One of the most popular approaches in condensed-matter physics is many-body perturbation theory (MBPT) based on Green's functions. Within the so called $GW$ approximation \cite{hedin65} to electron correlation, MBPT has become, over the last two decades, the method of choice for the calculations of quasiparticle band structures \cite{Aulbur,Gunnarsson,vidal2010,zunger2011,chulkov2013,gonze2013} and direct and inverse photo-emission spectra \cite{gatti-V2O3,Kotani07,gatti-VO2,Faleev_PRL2004,Rodl_PRB09,zunger2012} of many materials improving substantially over the results provided by static mean-field electronic structure methods. However $GW$ suffers from some fundamental shortcomings,\cite{Godby,pina09,vanSchilfgaarde06,Stan09,Caruso12,vonFriesen09} and, in particular, it does not capture strong correlation, unless one treats the system in a magnetically ordered phase. 
In particular, many paramagnetic insulators cannot be described correctly within $GW$. A paradigmatic example is the case of paramagnetic NiO, which is predicted to be a metal by $GW$.
Therefore, one has to go beyond simple approximations to the self-energy \cite{Aryasetiawan-Tmatrix,chulkov2004,romaniello_PRB12, guzzo_prl,guzzo2014,louie2013,kresse2014,Vollhardt2007} or explore novel routes to calculate Green's functions.\cite{lani_njp,berger_njp}  In this context, promising results have been reported for model systems by expressing the one-body Green's function as a continued fraction \cite{matho} as well as for solids \cite{Sharma13} using reduced density-matrix functional theory (RDMFT).\cite{Gilbert75} The RDMFT framework allows for the calculation of all the ground-state expectation values as functionals of the one-body reduced density matrix (1-RDM), provided that the functional is known. This, however, is in general not the case. In particular for spectral functions approximations have to be used. 

In this work we derive an expression for the spectral function, which is related to photoemission spectra, in terms of reduced density matrices (RDMs).
We show that simple approximations, which require the knowledge of the lowest $n$-body reduced density matrices ($n$-RDMs) only, can provide accurate photoemission spectra in model systems for moderate as well as for strong electron correlation. Our method overcomes the main problem of mean-field theories and $GW$ in correlated solids: as we show with the example of NiO, it correctly predicts this material to be insulating in both the antiferromagnetic and paramagnetic phases. 
The paper is organised as follows. In Sec.~\ref{Sec:Theory} we derive a new expression for the spectral function in terms of density matrices, we discuss simple approximations to it, and their physical meaning. In Sec.~\ref{Sec:Results} we illustrate, with the Hubbard model and the more realistic example of bulk NiO, how these approximations perform. Finally, in Sec.~\ref{Sec:Conclusions} we draw our conclusions and perspectives. 

\section{Theory\label{Sec:Theory}}
We start from the spectral representation of the time-ordered Green's function $G$ at zero temperature, which reads
\begin{equation}
G_{ij}(\omega)=\sum_k \frac{B_{ij}^{k,R}}{\omega-\epsilon_k^{R}-\I\eta}+\sum_k \frac{B_{ij}^{k,A}}{\omega-\epsilon_k^{A}+\I\eta},
\label{Eqn:SR_G}
\end{equation}
where $\epsilon_k^{R}=E_0-E_k^{N-1}$, $\epsilon_k^{A}=E_k^{N+1}-E_0$, $B^{k,R}_{ij}=\langle\Psi_0|\hat{c}_j^\dagger|\Psi_k^{N-1}\rangle\langle \Psi_k^{N-1}|\hat{c}_i|\Psi_0\rangle$, $B^{k,A}_{ij}=\langle\Psi_0|\hat{c}_i|\Psi_k^{N+1}\rangle\langle \Psi_k^{N+1}|\hat{c}_j^\dagger|\Psi_0\rangle$,  with $E_0$ and $\Psi_0$ the ground-state energy and wavefunction of the $N$-electron system and $E^{N\pm1}_k$ and $\Psi^{N\pm1}_k$ the $k$-th state energy and wavefunction of the $(N\pm1)$-electron system.
The superscripts `R' and `A' in \eqref{Eqn:SR_G} indicate the removal and addition parts of $G$, respectively. 
In the following we concentrate on the diagonal elements of $G$, which are related to photoemission spectra.  We choose to work in the basis set of natural orbitals $\phi_i$, i.e, the orbitals which diagonalize the 1-RDM, $\gamma(\mathbf{x},\mathbf{x}')=\sum_i n_i\phi_i(\mathbf{x})\phi^*_i(\mathbf{x}')$, where $0\leq n_i \leq 1$ are the occupation numbers and $\mathbf{x}=(\mathbf{r}, s)$ is a combined space-spin coordinate. {\color{black}Note that natural orbitals with $n_i=0$ are not uniquely defined. We fix them by assuming they correspond to the non-interacting solution.}  
In this basis set $\sum_k B_{ii}^{k,R}=n_i$ and $\sum_k B_{ii}^{k,A}=(1-n_i)$. 
Inspired by the \textit{numerical} effective-energy technique introduced in Refs \cite{berger_PRB2012,berger_PRB2012_2} that was designed to speed up convergence of a given spectral sum in the independent-particle framework, we present here a many-body effective-energy theory (MEET) to derive new \textit{expressions} for the many-body spectral functions in terms of RDMs. A separate treatment of removal and addition spaces turns out to be crucial. Let us first concentrate on the removal part. We define the effective energy $\delta^R_{i}(\omega)$ by
\begin{equation}
G_{ii}^R(\omega)=\sum_k \frac{B_{ii}^{k,R}}{\omega-\epsilon_k^{R}}= \frac{\sum_kB_{ii}^{k,R}}{\omega-\delta^R_i(\omega)}=\frac{n_i}{\omega-\delta^R_i(\omega)}.
\label{Eqn:EET}
\end{equation}
The effective energy $\delta^R_{i}(\omega)$ accounts for all the poles of the removal part of $G_{ii}$, which is in principle possible since it is frequency dependent. We now rewrite Eq.\ \eqref{Eqn:EET} as
\begin{align}
\delta^R_i(\omega)
=&\,\,\frac{\tilde{G}^R_{ii}(\omega)}{G^R_{ii}(\omega)},
\label{Eqn:delta_tot}
\end{align}
where
\begin{equation}
\tilde{G}^R_{ii}(\omega)=\sum_k\frac{\langle\Psi_0|\hat{c}_i^\dagger|\Psi_k^{N-1}\rangle\langle \Psi_k^{N-1}|[\hat{c}_i,\hat{H}]|\Psi_0\rangle}{\omega-\epsilon_k^{R}}.
\end{equation}
We can now introduce another effective energy $\tilde{\delta}^{R}_i(\omega)$ that accounts for all the poles of $\tilde{G}^R_{ii}(\omega)$. Working out the equations one arrives at $\tilde{\delta}^{R}_i(\omega)=\tilde{\tilde{G}}^R_{ii}(\omega)/\tilde{G}^R_{ii}(\omega)$, with 
\begin{equation}
 {\tilde{\tilde{G}}}^R_{ii}(\omega)=\sum_k\frac{\langle\Psi_0|[\hat{H},\hat{c}_i^\dagger]|\Psi_k^{N-1}\rangle\langle \Psi_k^{N-1}|[\hat{c}_i,\hat{H}]|\Psi_0\rangle}{\omega-\epsilon_k^{R}}.
\end{equation}
This leads to
\begin{equation}
\delta^R_{i}(\omega)=\frac{\frac{\tilde{n}^R_{i}}{\omega-\tilde{\delta}^R_{i}(\omega)}}{\frac{n_{i}}{\omega-\delta^R_{i}(\omega)}}=\frac{\tilde{n}^R_{i}}{n_{i}}\frac{\omega-\frac{\tilde{G}^{R}_{ii}(\omega)}{G^{R}_{ii}(\omega)}}{\omega-\frac{\tilde{\tilde{G}}^{R}_{ii}(\omega)}{\tilde{G}^{R}_{ii}(\omega)}}.
\label{eqn:CapEET_delta_tot}
\end{equation}
In principle, one could continue this procedure \textit{ad infinitum}. 
In practice, however, one has to truncate the series. This can be done in various ways. 
Here we choose a truncation that guarantees the exact results for the Hubbard dimer \tobe{at 1/2 filling} at all orders, as we will discuss \tobe{later}. This is obtained by assuming that at a certain order the poles of $G_{ii}^R$, $\tilde{G}_{ii}^R$, ..., expressed in terms of the respective effective energies $\delta^R_i$, $\tilde{\delta}^R_i$, ..., are the same.
The first two approximations to $\delta^R_i(\omega)$ read
\begin{equation}
\delta^{R,(1)}_i=\frac{\tilde{n}^R_i}{n_i},
\label{delta1}
\end{equation}
\begin{equation}
\delta^{R,(2)}_i(\omega)=\frac{\tilde{n}^R_{i}}{n_{i}}\frac{\omega-\frac{\tilde{n}^R_{i}}{n_{i}}}{\omega-\frac{\tilde{\tilde{n}}^R_{i}}{\tilde{n}^R_{i}}},
\label{delta2R}
\end{equation}  
where \noeqref{delta2R}
\begin{align}
\tilde{n}^R_{i}&=
\langle\Psi_0|\hat{c}_i^\dagger[\hat{c}_i,\hat{H}]
|\Psi_0\rangle=h_{ii} n_i+\sum_{jkl}V_{ijkl}\Gamma^{(2)}_{klji},
\label{ntilde}
\end{align}
with $\Gamma^{(2)}_{klji}=\langle\Psi_0|\hat{c}^\dagger_i\hat{c}^\dagger_j\hat{c}_l\hat{c}_k|\Psi_0\rangle$ the matrix elements of the 2-RDM.
Here $h_{ij}=\int d\mathbf{x}\phi_i^*(\mathbf{x})h(\mathbf{r})\phi_j(\mathbf{x})$ are the matrix elements of the one-particle noninteracting Hamiltonian $h(\mathbf{r})=-\nabla^2/2+v_{\text{ext}}(\mathbf{r})$, and $V_{ijkl}=\int d\mathbf{x}d\mathbf{x}'\phi^*_i(\mathbf{x})\phi^*_j(\mathbf{x}')v_c(\mathbf{r},\mathbf{r}')\phi_k(\mathbf{x})\phi_l(\mathbf{x}')$ are the matrix elements of the Coulomb interaction $v_c$. The expression of $\tilde{\tilde{n}}^R_{i}=\langle\Psi_0|[\hat{H},\hat{c}_i^\dagger][\hat{c}_i,\hat{H}]
|\Psi_0\rangle$ in terms of 1-, 2, and 3-RDMs is given in App.\ \ref{App:1}. 

Similarly for the addition energies one can introduce an effective energy $\delta^A_{i}(\omega)$, and derive approximations. The first two approximations to $\delta^A_{i}(\omega)$ read
\begin{align}
\delta^{A,(1)}_i&=\frac{\tilde{n}^A_i}{1-n_i},
\label{delta1_A}\\
\delta^{A,(2)}_i(\omega)&=\frac{\tilde{n}^A_{i}}{1-n_{i}}\frac{\omega-\frac{\tilde{n}^A_{i}}{1-n_{i}}}{\omega-\frac{\tilde{\tilde{n}}^A_{i}}{\tilde{n}^A_{i}}},
\label{delta2A}
\end{align}
where \noeqref{delta2A}\noeqref{delta1_A}
\begin{equation}
\tilde{n}^A_{i}=
\langle\Psi_0|[\hat{c}_i,\hat{H}]\hat{c}_i^\dagger
|\Psi_0\rangle=
h_{ii}+\sum_j\left(V_{ijij}-V_{ijji}\right)n_j-\tilde{n}^R_{i},
\label{ntilde_A}
\end{equation}
and the expression of $\tilde{\tilde{n}}^A_{i}=\langle\Psi_0|[\hat{c}_i,\hat{H}][\hat{H},\hat{c}_i^\dagger]
|\Psi_0\rangle$ in terms of 1-, 2-, and 3-RDMs is given in App.\ \ref{App:1}. 

{\color{black}An important point to note is that, interestingly, for a given natural orbital $\phi_i$, removal and addition effective energies are different: this is essential to open a gap, as we will illustrate in the Hubbard model.}

The spectral function can then be written as
\begin{equation}
A_{ii}(\omega)=n_i\delta(\omega-\delta^R_i(\omega))+(1-n_i)\delta(\omega-\delta^A_i(\omega)),
\label{Eqn:SF}
\end{equation}
which satisfies the sum rule $\int^{\infty}_{-\infty}d\omega A_{ii}(\omega)=1$. Starting from $\delta^{(2)}$ one could, in principle, get complex poles because the equations become nonlinear in the frequency. In this case the sum rule is not satisfied. However, such complex poles do not occur for $\delta^{(2)}$, at least in the model systems we studied. For higher-order approximations remedies
such as the regularization of unphysical poles could be envisaged \footnote{See Ref.\ \onlinecite{berger_PRB2012_2} for an example in the case of independent particles.}. Our goal here is to derive simple and physically-motivated expressions for the spectral function, and this is obtained using $\delta^{(1)}$ and $\delta^{(2)}$, as we will show later.

The expression (\ref{Eqn:SF}) looks similar to the one reported in Ref.\ \onlinecite{Sharma13} in the context of RDMFT. In that case (referred to as DER in the following), however, removal and addition energies are calculated in a different way, namely as functional derivatives of the ground-state total energy with respect to the occupation numbers. {\color{black}
Our method, instead, is not bound to RDMFT: it can be used with any approach which can provide one with RDMs. It does therefore not require a total energy that is a functional of occupation numbers. For example the RDMs could be obtained from quantum Monte Carlo (see, e.g., Refs           
\onlinecite{Alavi_JCP,Alavi_nature}).

\begin{figure}
\begin{center}
\includegraphics[width=0.45\textwidth]{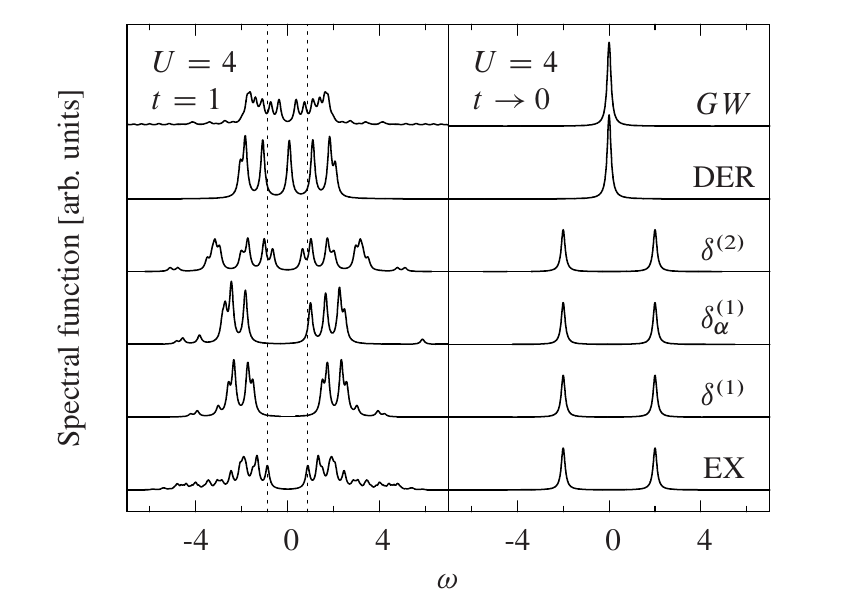}
\end{center}
\vspace{-18pt}
\caption{Spectral function for a 12-site Hubbard ring at 1/2 filling: exact (EX) \textit{vs} MEET with exact RDMs ($\delta^{(1)}$ and $\delta^{(2)}$), MEET with approximate RDMs ($\delta_{\alpha}^{(1)}$, $\alpha=0.5$), DER method (with $\alpha=0.5$) and $GW$ method. Peaks are broadened with a Lorentzian of width $\eta=0.1$.}
\label{figure3}
\end{figure}

The physical meaning of $\delta_i^{R,(1)}$ can be seen by combining Eqs \eqref{delta1} and \eqref{ntilde}. 
This gives a weighted average of all removal poles of $G^R_{ii}$, since one gets $\delta_i^{R,(1)}=\sum_k B_{ii}^{k,R} \epsilon_k^R / \sum_k B_{ii}^{k,R}$. In other words $\delta^{R,(1)}_{i}$ is equal to the first moment of $G^R_{ii}(\omega)$. Here the $n$-th moment is defined as $\mu^R_{n,i}=\sum_k B_{ii}^{k,R} (\epsilon_k^R)^n/ \sum_k B_{ii}^{k,R}$. A similar relation can be derived for $\delta_i^{A,(1)}$. Moreover the first and second moments of the approximate Green's function generated by $\delta_i^{R/A,(2)}$ are equal to the first and  second moments of the exact Green's function. For $\delta_i^{R/A,(n)}$ with $n>2$ higher moments are involved.\footnote{We verified analytically that similar expressions  are satisfied for $n>2$, at least up to $n=4$.}
Thanks to the fact that the first moment $\mu_{1,i}^R$ of the approximate Green's function is equal to the exact one, 
the total energy calculated using the one-body Green's function is exact, provided that the exact RDMs are used. 
Using the Galitskii-Migdal equation, one can indeed express the exact total energy in terms of $\mu^R_{1,i}$ and $n_i$ as $
E_0= \sum_i n_i \left(\mu^R_{1,i} + h_{ii}\right)/2$.

\begin{figure}
\begin{center}
\includegraphics[width=0.45\textwidth]{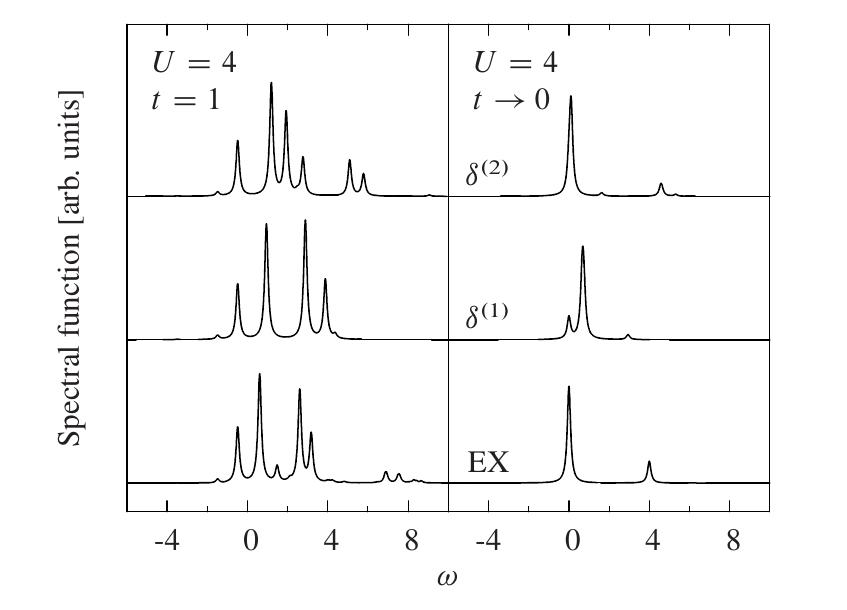}
\end{center}
\vspace{-18pt}
\caption{Spectral function for a 6-site Hubbard ring at 1/6 filling: exact (EX) \textit{vs} MEET with exact RDMs ($\delta^{(1)}$ and $\delta^{(2)}$). Peaks are broadened with a Lorentzian of width $\eta=0.1$.}
\label{figure2}
\end{figure}

\section{Results\label{Sec:Results}}
In this section we show how the approximations derived above work in practice. 
\subsection{Hubbard model}
First we use the Hubbard model as a test system. The important parameters of the model are the  on-site Coulomb interaction $U$ and the hopping term $-t$. For the Hubbard dimer at 1/2 filling the method is exact for all $\delta^{(n)}$ with $n\geq 1$. 
The case with more sites at 1/2 filling is highly non trivial. {\color{black} For small rings, the simple approximation $\delta^{(1)}$ suffices to give an accurate spectrum at all interaction strengths, although it tends to overestimate the band gap. When we consider larger rings, at the level of $\delta^{(1)}$, the spectral shape is still good, but the overestimation of the band gap is more evident, as can be seen in Fig.\ \ref{figure3} (left panel), where we present the case of a 12-site ring. Furthermore we find that, for a fixed interaction $U$, the ratio between the $\delta^{(1)}$ band gap and the exact band gap increases going from the 1D to the 2D infinite Hubbard model (see App.~\ref{App:2}); this suggests that the error of $\delta^{(1)}$ in reproducing the band gap increases with the dimensionality. One has to go to $\delta^{(2)}$ to partially correct this overestimation.  

Also away from 1/2 filling one has to go to $\delta^{(2)}$ to have overall better results, in particular in the atomic limit (see Fig.\ \ref{figure2}).
Note that, although the spectral profile given by $\delta^{(1)}$ is in good agreement with the exact spectrum at $t=1$, as shown in Fig.\ \ref{figure2}, the analysis of the peaks in the energy range $0<\omega<5$ has revealed a mixed quasiparticle/satellite character. 
Moreover, in the atomic limit, the main peak at $\omega=0$ in the exact spectrum is a superposition of a removal peak and an addition peak.
In this limit, $\delta^{(1)}$ opens a band gap around $\omega=0$, which is not present in the exact results.
Using $\delta^{(2)}$ tends to correct these errors. This indicates that $\delta^{(1)}$ is not a good approximation in metallic systems.

\begin{figure*}
\begin{center}
\includegraphics[width=0.45\textwidth]{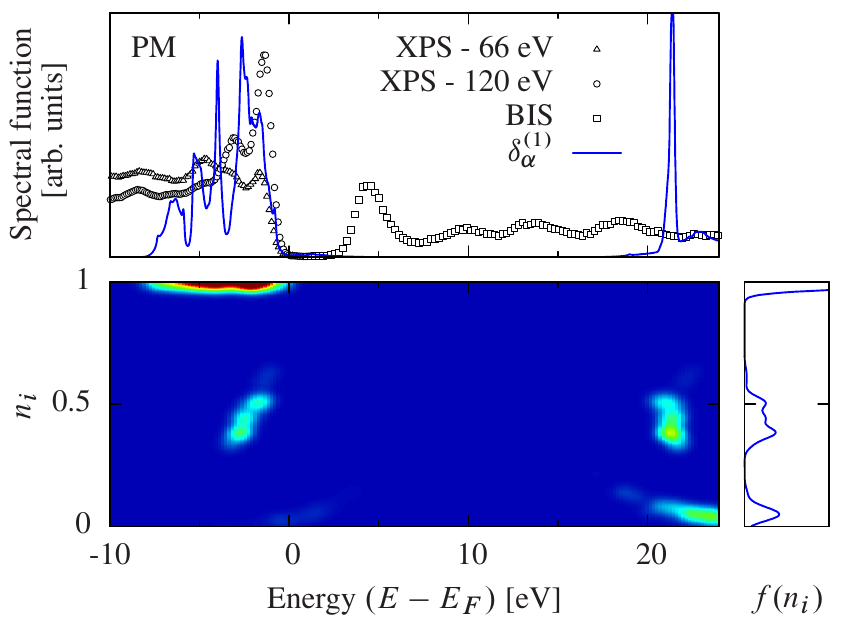}\hspace{10pt}
\includegraphics[width=0.45\textwidth]{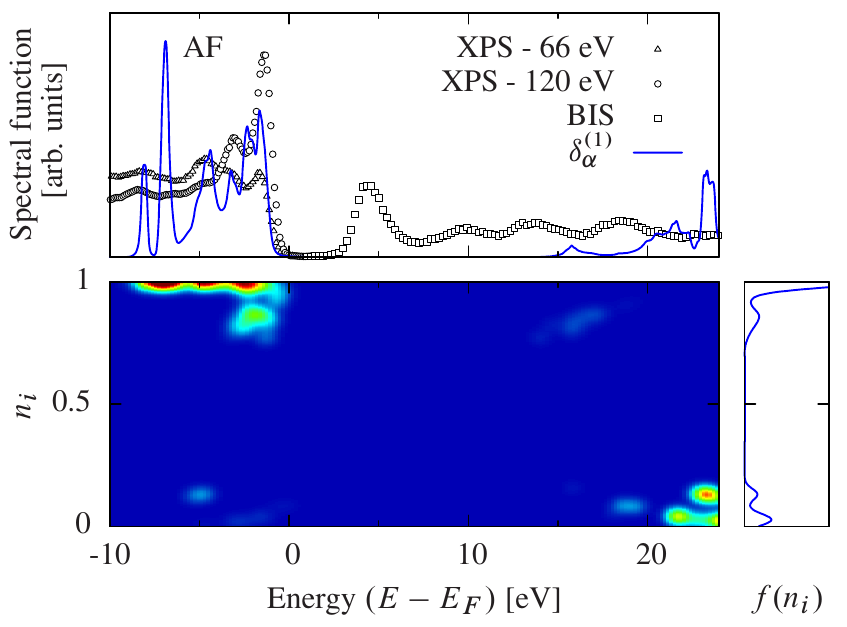}
\end{center}
\vspace{-15pt}
\caption{(Color online) Paramagnetic (left panel) and antiferromagnetic (right panel) bulk NiO: experimental photoemission spectrum \cite{PhysRevLett.53.2339} \textit{vs} MEET spectrum ($\delta^{(1)}_{\alpha}$, $\alpha$=0.65). The color map and the distribution $f(n_i)$
illustrate the occupation numbers $n_i$ which play a role into the spectrum for the reported energy range. }
\label{figure5}
\end{figure*}

There are two striking features of the results obtained with our method: (i) there are satellites, even with a static approximation ($\delta^{(1)}$), i.e., more energies than the number of natural orbitals; (ii) there is a gap in the atomic limit 
(see right panel of Fig.\ \ref{figure3})
without breaking the symmetry of the system (i.e., without localizing the spins each on a site).  The first feature can be understood looking at the spectral weights in the spectral function \eqref{Eqn:SF}, which are $n_i$ for the removal energies and $1-n_i$ for the addition energies. 
As long as the occupation numbers are 0 or 1, as in the noninteracting case, for each $n_i$ one sees either a removal or an addition peak. When, instead, $0<n_i<1$, then, for each $n_i$ one gets both a removal and an addition peak. In other words, at the level of $\delta^{(1)}$, for each orbital one gets two energies. These two energies are related by:
\begin{equation}
\delta^{A,(1)}_i=\delta^{R,(1)}_i-\frac{1}{n_i(1-n_i)}\sum_{jkl}V_{ijkl}\Gamma^{(2)}_{c,klji},
\label{Eqn:deltaA-R}
\end{equation}
where $\Gamma^{(2)}_{c}$ is the correlation contribution to the 2-RDM  (see App.~\ref{App:1}). 
{\color{black}In the Hubbard dimer at 1/2 filling, for example, $\sum_{jkl}V_{ijkl}\Gamma^{(2)}_{c,klji}=-U/2\sqrt{n_bn_a}$, with $n_a$ and $n_b$ the occupation numbers of the bonding and antibonding natural orbitals, respectively. In the atomic limit the occupation numbers tend to 1/2, and therefore the spectral function consists of one removal and one addition energy peak of equal weight, each being a superposition of the bonding and the antibonding component; the band gap is hence between two peaks of the same component, and it equals $U$, as given by Eq.\ \eqref{Eqn:deltaA-R}. The $GW$ approximation, instead, can open a gap only between different components, i.e., it describes a ``quasiparticle" gap; it hence fails to describe the ``correlation" gap in the the Hubbard dimer in the atomic limit.}

Note that, when the Hartree-Fock (HF) approximation is used for the 2-RDM, the effective energies $\delta_i^{R/A,(1)}$ are equal to
the removal/addition energies obtained with the HF self-energy. However, since $n_i=1$ or $0$, only one of the two appears in the spectral function.

In real situations the exact RDMs are not known. Here we focus on $\delta^{(1)}$, since it requires only the knowledge of the 1- and 2-RDMs, which can be obtained within RDMFT. In this framework the 2-RDM is a functional of the 1-RDM. This functional is not known, but approximations
are available. In this work we will use the power functional $\Gamma^{(2)}(\mathbf{x},\mathbf{x}^\prime;\mathbf{x},\mathbf{x}^\prime)\approx\gamma(\mathbf{x}, 
  \mathbf{x})\gamma(\mathbf{x}^\prime, \mathbf{x}^\prime)
-\gamma^\alpha(\mathbf{x}, 
  \mathbf{x}^\prime)\gamma^{\alpha}(\mathbf{x}^\prime, \mathbf{x})$ where
$ \gamma^\alpha(\mathbf{x},\mathbf{x}^\prime)=\sum_{i} n^\alpha_{i}\phi_{i}(\mathbf{x})\phi^*_i(\mathbf{x}^\prime)$, with $0.5\leq \alpha\leq 1$.\cite{sharma_PRB08}

First we test this approximation on the Hubbard model. In Fig.\ \ref{figure3} we report the results for a 12-site ring using $\alpha=0.5$. We obtain good results (although the power functional does not recover the particle-hole symmetry) and, in particular, the band gap opens in the atomic limit, without breaking the symmetry.\cite{stefano_JCP2015} We note that the gap strongly depends on the value of $\alpha$: it has its maximum value for $\alpha=0.5$ while it disappears for $\alpha=1$ which corresponds to HF.
For comparison we also report the results obtained using $GW$ and the DER method: in the atomic limit these methods do not open any gap, unless the symmetry of the system is broken.\cite{stefano_JCP2015} 

This raises the question whether the MEET, using the simple approximations that are successful in the models, could also open the band gap of a real Mott insulator.

\subsection{Realistic systems: the example of NiO}
We implemented our approach in a modified version of the full-potential linearized augmented plane wave (FP-LAPW) code Elk,\cite{elk} {\color{black}with practical details of the calculations following the scheme described in Ref.\ \onlinecite{sharma_PRB08}.} We apply the method to bulk NiO, which is a prototypical strongly correlated material. This system shows antiferromagnetic behavior below the N\'eel temperature, and the photoemission spectrum is similar for the paramagnetic and the antiferromagnetic phases,\cite{PhysRevB.54.10245} with a band gap of about 4.3 eV.\cite{PhysRevLett.53.2339} Already at the level of LDA, the antiferromagnetic phase shows a band gap, although too small, due to quasiparticle splitting.
The spectrum in this phase can be well described by $GW$ \cite{Faleev_PRL2004, Rodl_PRB09} as well as RDMFT using the DER method.\cite{Sharma13} However, the real challenge is to open a gap without symmetry breaking (see App.\ \ref{App:3}) as it should happen in the paramagnetic phase. To this aim we model the paramagnetic phase as nonmagnetic. In this case the gap is purely due to correlation,
which LDA and $GW$ fail to describe. Our approach, instead, opens a gap in both phases.} This is shown in Fig.\ \ref{figure5}. Note that our results for both phases are compared with experiment for the antiferromagnetic phase.\cite{PhysRevLett.53.2339} This comparison is meaningful since
the observed phothoemission spectrum of NiO is almost unaffected by the magnetic phase transition.\cite{PhysRevB.54.10245} For the calculations we used the experimental lattice constants and the power functional ($\alpha=0.65$) with the self-interaction correction proposed by Goedecker and
Umrigar.\cite{GU_PRL98} From the analysis of the
occupation numbers it emerges that the physics underlying the band gap opening in the two phases is indeed different: in the antiferromagnetic case it is mainly due to occupation numbers close to one or zero, whereas in the paramagnetic
phase it is mainly due to occupation numbers around 0.5.  This is in line with an analogous analysis on the Hubbard dimer.\cite{stefano_JCP2015} It still remains to improve the band gap, which is overestimated in our method. This finding is consistent with the results on large Hubbard rings, which indicate that this overestimation is due to the use of $\delta^{(1)}$, and that the use of  $\delta^{(2)}$ might improve the spectrum and the band gap. This would require approximations for the 3-RDM, which is beyond the scope of this paper. However, 
we can get a rough estimation of the influence of $\delta^{(2)}$ on the spectrum of NiO by using an effective $\delta^{(1)}$ in which some of the effects due to $\delta^{(2)}$ are included. Using the 12-site Hubbard chain (for which both $\delta^{(1)}$ and $\delta^{(2)}$ can be calculated exactly) we calculated the screening of $\delta^{(1)}$ that reproduces the effect of $\delta^{(2)}$. The screening (which we call here $\beta$) depends on the component of $G$ (i.e., on the natural orbital) one is looking at, and has been defined according to 
\begin{align}
\delta_i^{R,(1)} &=h_{ii} +\sum_{j}V_{ijij} n_j+ \frac{\beta_i}{n_i} \sum_{jkl}V_{ijkl}\Gamma^{(2)}_{\text{xc},klji},
\label{ntilde}
\end{align}
where $\Gamma^{(2)}_{\text{xc},klji}=\Gamma^{(2)}_{klji}-n_i n_j \delta_{ik}\delta_{jl}$ is the exchange-correlation part of the 2-RDM (see App.~\ref{App:1}).
For most of the natural orbitals that are responsible for the gap in the Hubbard model, the values of $\beta_i$ are in the range $0<\beta_i<1$. 
Therefore, for the calculation of the spectrum of NiO, we choose $\beta_i$ between 0 and 1 and assume it to be the same for each orbital. This leads to a reduction of the band gap;
in particular, with $\beta=0.8$ the spectral function of paramagnetic NiO is in better agreement with  experiment, as illustrated in Fig.\ \ref{figure8}. Further decreasing $\beta$ can eventually close the gap.
These findings indicate that one could envisage to include in an approximate way higher-order terms using a proper terminating function.

\begin{figure}[b]
\begin{center}
\includegraphics[width=0.45\textwidth]{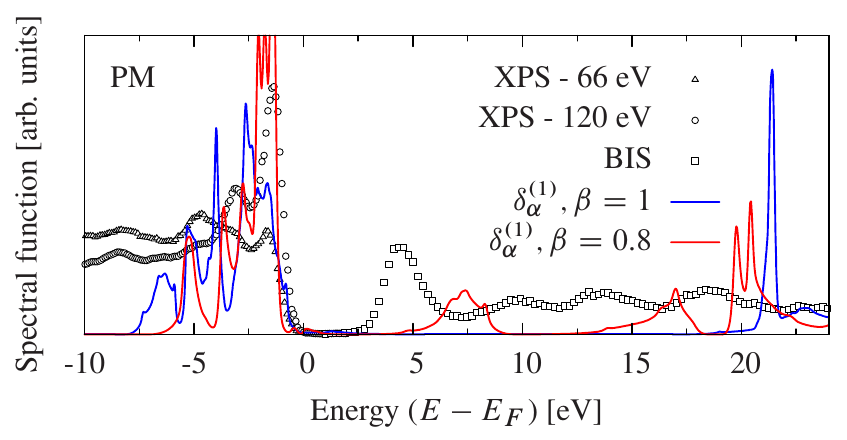}
\end{center}
\vspace{-15pt}
\caption{Paramagnetic bulk NiO: experimental photoemission spectrum \cite{PhysRevLett.53.2339}  \textit{vs} MEET spectrum ($\delta_\alpha^{(1)}$, with the power functional, $\alpha=0.65$, and without screening ($\beta=1$) and with screening $\beta=0.8$).}
\label{figure8}
\end{figure}

\section{Conclusions and prospectives\label{Sec:Conclusions}}
In conclusion, we have derived an expression for the spectral function in terms of RDMs.
Simple approximations can give accurate spectra for finite model systems in the weak as well as in the strong correlation regime. In particular the method correctly reproduces the atomic limit of Hubbard systems without breaking the symmetry of the system. We applied a simple approximation depending only on the 1- and 2-RDMs to bulk NiO within the computationally efficient RDMFT. Our method produces qualitatively correct photoemission spectra for the antiferromagnetic and paramagnetic phases, although the band gap is  overestimated. 
The study of the Hubbard model indicates how this problem might be overcome in the future.
Our method indicates a promising way to approach the problem of band gaps due to correlation effects in a relatively simple manner.

\section{Acknowledgments}
The research leading to these results has received funding
from the European Research Council under the European
Union's Seventh Framework Programme (FP/2007-2013) / ERC Grant Agreement n.~320971.
Discussion within the Collaboration Team on Correlation of the European Theoretical Spectroscopy facility (ETSF) is greatly acknowledged. 
S.D.S. and P.R. thank M. Gatti, M. Guzzo, and S. Sharma for fruitful discussions.
\appendix
\section{Approximations to $\delta_{i}^{R/A}$ in terms of reduced density matrices\label{App:1}}
In the following we express $\delta_{i}^{R/A,(n)}$, with $n=1,2$ in terms of RDMs.
Let us consider the following many-body Hamiltonian
\begin{equation}
 \hat{H} =\hat{H}_0+\hat{V}=\sum_{i} h_{ij}\hat{c}^{\dagger}_{i} \hat{c}_{j} + 
\frac{1}{2} \sum_{ijkl} V_{ijkl} \hat{c}^\dagger_{i}\hat{c}^\dagger_{j}\hat{c}_{l}\hat{c}_{k} 
\label{eqn:CapMEET_Hubbard_H}
\end{equation}
where $\hat{c}^\dagger_{i}$ and $\hat{c}_{i}$ are the creation and annihilation operators in the basis of natural orbitals $\phi_{i}(\vect{x})$.
Here $h_{ij}=\int d\vect{x}\phi_i^*(\vect{x})h(\vect{r})\phi_j(\vect{x})$ are the matrix elements of the one-particle noninteracting Hamiltonian $h(\vect{r})=-\nabla^2/2+v_{\text{ext}}(\vect{r})$, and 
$V_{ijkl}=\int d\vect{x}d\vect{x}'\phi^*_i(\vect{x})\phi^*_j(\vect{x}')v_c(\vect{r},\vect{r}')\phi_k(\vect{x})\phi_l(\vect{x}')$,
are the matrix elements of the Coulomb interaction $v_c$. 
Using the Hamiltonian \eqref{eqn:CapMEET_Hubbard_H}, we can evaluate the commutators appearing in the expressions for $\tilde{n}_i^{R/A}$ and 
$\tilde{\tilde{n}}_i^{R/A}$.
We obtain the following relations
\begin{equation}
\tilde{n}_i^R=h_{ii}n_i+\sum_{jkl}V_{ijkl}\Gamma_{klji}^{(2)},
\label{eqn:ChapMEET_ntildeRGamma}
\end{equation}
\begin{align}
\tilde{\tilde{n}}^R_i=&\,\,h_{ii}^2 n_i+h_{ii}\sum_{jkl}\left(V_{ijkl}\Gamma^{(2)}_{klji}+V_{jkil}\Gamma^{(2)}_{ilkj}\right)\nonumber\\
&+\sum_{jklk'l'}V_{jkil}V_{ilk'l'}\Gamma^{(2)}_{k'l'kj}\nonumber\\
&+\sum_{jklj'k'l'}V_{jkil}V_{ij'k'l'}\Gamma^{(3)}_{k'l'lj'kj},
\label{eqn:ChapMEET_ntildetildeRGamma}
\end{align}
\begin{equation}
\tilde{n}_i^A=h_{ii}(1-n_i)+\sum_{j}\left(V_{ijij}-
 V_{ijji}\right)n_{j} 
  -\sum_{jkl}V_{ijkl}
  \Gamma_{klji}^{(2)},
  \label{eqn:ChapMEET_ntildeAGamma}
\end{equation}
\begin{align}
\tilde{\tilde{n}}^A_i=&\,\,h_{ii}^2(1-n_i)
+2 h_{ii}\sum_j  (V_{ijij}-V_{ijji})n_j\nonumber\\
&+h_{ii}\sum_{jkl}\left(V_{ijkl}\Gamma^{(2)}_{klij}+V_{jkil}\Gamma^{(2)}_{likj}\right)\nonumber\\
&+\sum_{jkl} \left(V_{iljk}-V_{ilkj}\right)V_{jkil}n_l\\
&+\sum_{jklj'k'}\left(V_{ij'k'j}-V_{ij'jk'}\right)\left(V_{jkil}-V_{kjil}\right)\Gamma^{(2)}_{lk'kj'} \nonumber\\
&+\sum_{jklj'k'l'}V_{ij'k'l'}V_{jkil}\Gamma^{(3)}_{lk'l'kjj'}.
\label{eqn:ChapMEET_ntildetildeAGamma}
\end{align}
where 
$
\Gamma^{(2)}_{ijkl}=\bra{\Psi_0}\hat{c}^\dagger_{l}\hat{c}^\dagger_{k}\hat{c}_{j}\hat{c}_{i}\ket{\Psi_0}
$
and 
$
\Gamma^{(3)}_{ijklmn}=\langle\Psi_0|\hat{c}^\dagger_{n}\hat{c}^\dagger_{m}\hat{c}^\dagger_{l}\hat{c}_{k}\hat{c}_{j}\hat{c}_{i}|\Psi_0\rangle
$
are the matrix elements of the 2-RDM and 3-RDM, respectively. Using Eqs \eqref{eqn:ChapMEET_ntildeRGamma}-\eqref{eqn:ChapMEET_ntildetildeAGamma} 
we get the expressions of $\delta_{i}^{R/A,(1)}$ and $\delta_{i}^{R/A,(2)}$ in terms of RDMs. Here for simplicity we give only the expressions of $\delta_{i}^{R/A,(1)}$; they read
\begin{align}
 \delta_i^{R,(1)}&=h_{ii}+\frac{1}{n_i}\sum_{jkl}V_{ijkl}\Gamma^{(2)}_{klji},\label{eqn:ChapMEET_deltaRGamma}\\
 \delta_i^{A,(1)}&=h_{ii}+\frac{1}{1-n_i}\left[\sum_{j}(V_{ijij}-V_{ijji})n_{j} 
  -\sum_{jkl}V_{ijkl}
  \Gamma_{klji}^{(2)}\right].
  \label{eqn:ChapMEET_deltaAGamma}
\end{align}
The 2-RDM can be explicitly decomposed in terms of Hartree and  exchange-correlation contributions, respectively, as $\Gamma^{(2)}_{klji}=n_i n_j \delta_{ik}\delta_{jl}+\Gamma^{(2)}_{\text{xc},klji}$, or in terms of Hartree, exchange, and the correlation contributions, respectively,  as $\Gamma^{(2)}_{klji}=n_i n_j \delta_{ik}\delta_{jl}-n_i n_j \delta_{il}\delta_{jk}+\Gamma^{(2)}_{c,klji}$. In this case 
Eqs \eqref{eqn:ChapMEET_deltaRGamma} and \eqref{eqn:ChapMEET_deltaAGamma} can be rewritten as
\begin{align}
 \delta_i^{R,(1)}=&\,\,h_{ii}+\sum_{j}V_{ijij}n_j+\frac{1}{n_i}\sum_{jkl}V_{ijkl}\Gamma^{(2)}_{\text{xc},klji},\label{eqn:ChapMEET_deltaRGamma_xc}\\
= &\,\,h_{ii}+\sum_{j}(V_{ijij}-V_{ijji})n_j+\frac{1}{n_i}\sum_{jkl}V_{ijkl}\Gamma^{(2)}_{c,klji}\label{eqn:ChapMEET_deltaRGamma_c}\\
 \delta_i^{A,(1)} = &\,\,h_{ii}+\sum_{j} V_{ijij}n_j \nonumber\\
 &-\frac{1}{1-n_i}\left[\sum_j V_{ijji}n_{j} 
   +\sum_{jkl}V_{ijkl}\Gamma_{\text{xc},klji}^{(2)}\right]
  \label{eqn:ChapMEET_deltaAGamma_xc}\\
  =&\,\,h_{ii}+\sum_{j}(V_{ijij}-V_{ijji})n_j
 -\frac{1}{1-n_i}\sum_{jkl}V_{ijkl}\Gamma_{c,klji}^{(2)}
  \label{eqn:ChapMEET_deltaAGamma_c}.
\end{align}

\noeqref{eqn:ChapMEET_ntildetildeRGamma}\noeqref{eqn:ChapMEET_ntildeAGamma}

\section{Performance of $\delta_{\alpha}^{(1)}$ in 1D and 2D infinite Hubbard models at half filling\label{App:2}}
\begin{figure}
\begin{center}
\includegraphics[width=0.45\textwidth]{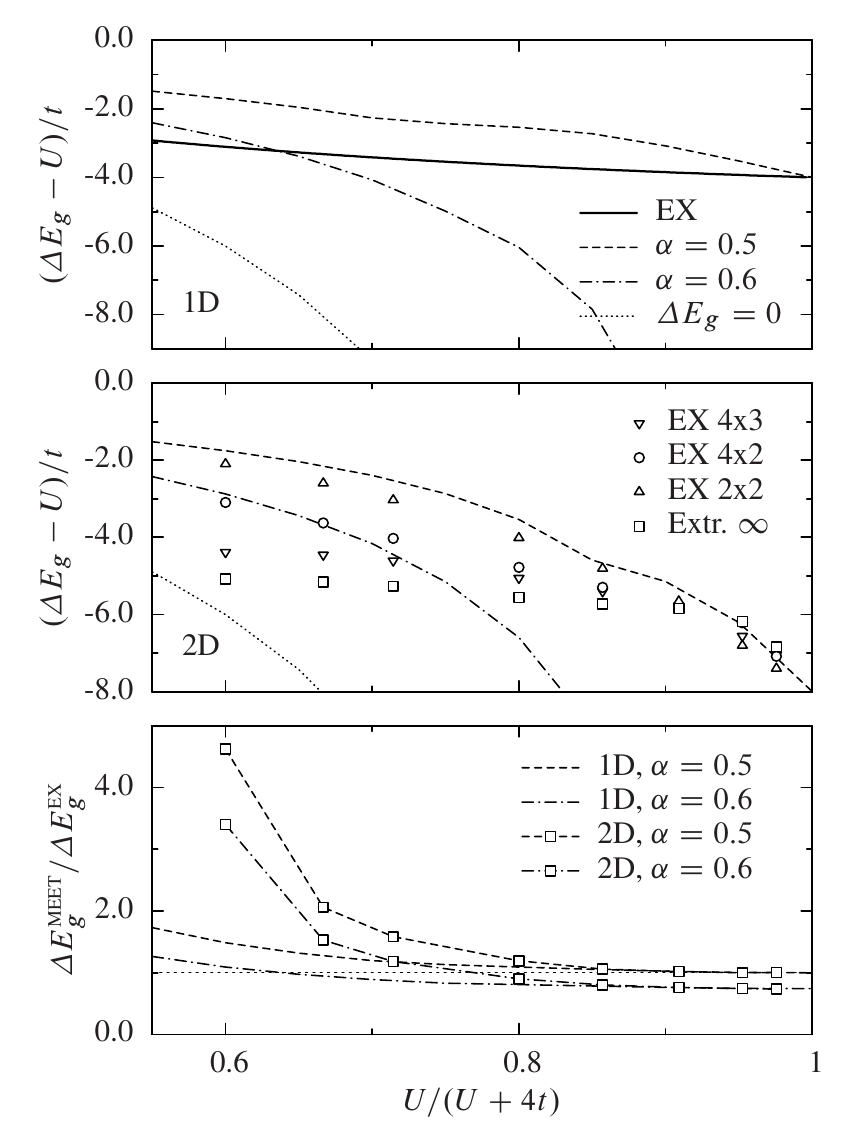}
\end{center}
\vspace{-10pt}
\caption{Deviation of the fundamental gap $\Delta E_g$ from $U$ as a function of $U/(U+4t)$ for the 1D (top panel) and 2D (middle panel) infinite Hubbard models at 1/2 filling. Dashed and dotted lines are obtained using the MEET method ($\delta^{(1)}_{\alpha}$, $\alpha=0.5, 0.6$). 
The black solid line in the top panel is the exact result derived from the Bethe-ansatz solution.\cite{Lieb68} The curve corresponding to $\Delta E_g=0$ is reported as reference.
The triangles, circles, and squares in the middle panel are finite-size exact calculations for the 2x2, 4x2, and 4x3 2D Hubbard clusters, respectively.
The squares are the extrapolated values to the infinite 2D system.
In the bottom panel the ratio between the $\delta^{(1)}_{\alpha}$ band gap and exact band gap ($\Delta E^{MEET}_g/\Delta E^{EX}_g$) is reported as function of $U/(U+4t)$. As a guide for the eye we reported $\Delta E^{MEET}_g/\Delta E^{EX}_g=1$ with a thin dashed line.}
\label{fig:GapHubbard}
\end{figure}
\begin{figure*}[t]
\begin{center}
\includegraphics[width=0.45\textwidth]{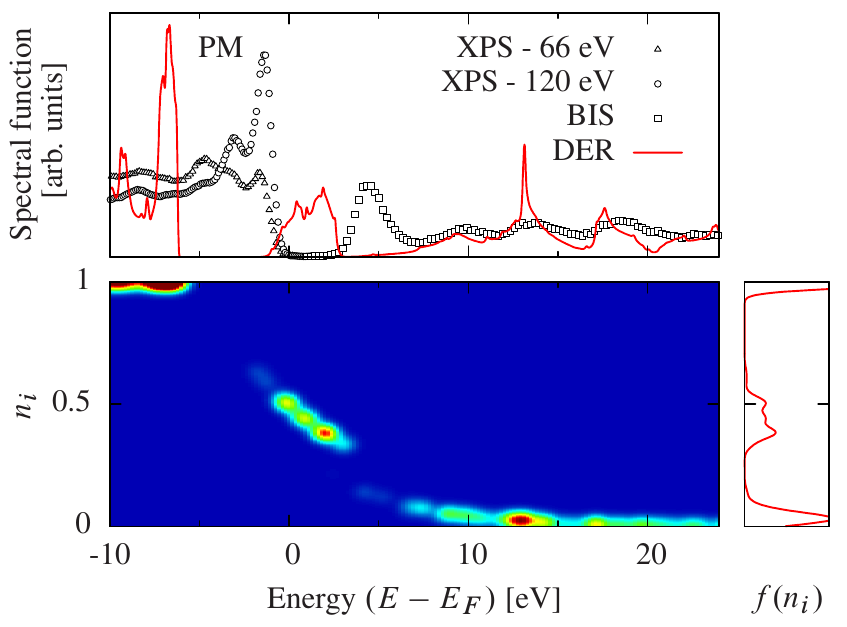}\hspace{10pt}
\includegraphics[width=0.45\textwidth]{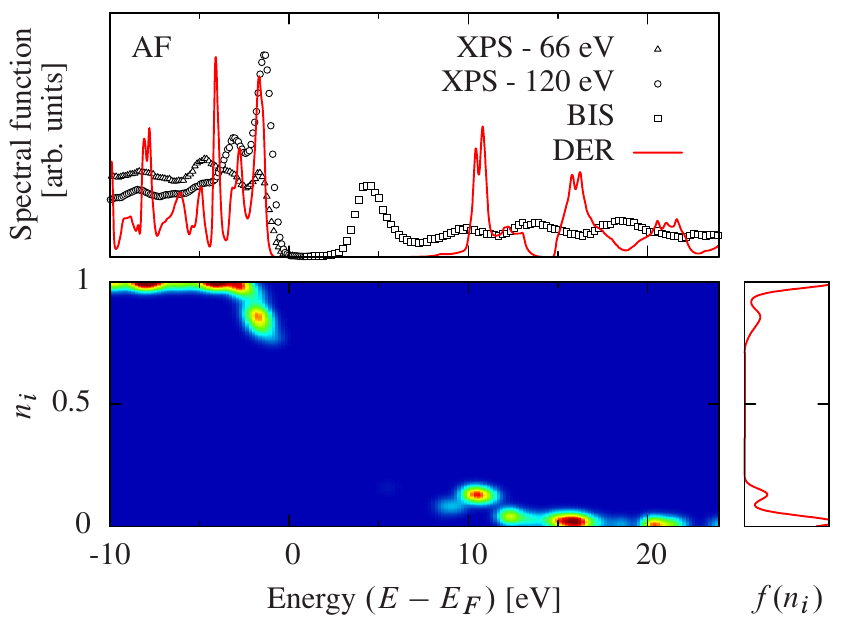}
\end{center}
\vspace{-15pt}
\caption{Paramagnetic (left panel) and antiferromagnetic (right panel) bulk NiO: experimental photoemission spectrum \cite{PhysRevLett.53.2339}  \textit{vs} DER spectrum (with the power functional, $\alpha$=0.65). The color map and the distribution $f(n_i)$
illustrate the occupation numbers $n_i$ which play a role into the spectrum for the reported energy range.}
\label{figure7}
\end{figure*}

In Fig.~\ref{fig:GapHubbard} (top and middle panels) the fundamental gap $\Delta E_g$, calculated using the MEET method ($\delta_{\alpha}^{(1)}$, $\alpha=0.5$, $0.6$), is reported as a function of the interaction $U$ for the 1D and 2D infinite Hubbard models at 1/2 filling.
For the 1D system the MEET results are compared to the exact Bethe-Ansatz solution,\cite{Lieb68} whereas for 2D they are compared to our exact results for finite-size 2D Hubbard clusters and the extrapolated results to the infinite 2D system. 
In the figure we report only the range of $U$ for which the 2D system is an insulator. 
For each value of $U$, $\delta^{(1)}_{\alpha}$ can produce the correct band gap by properly tuning $\alpha$. 

In the bottom panel of Fig.~\ref{fig:GapHubbard} we compare the ratio between the $\delta^{(1)}_{\alpha}$ band gap and the exact band gap ($\Delta E^{MEET}_g/\Delta E^{EX}_g$) for the two Hubbard models as a function of $U$: for a fixed $U$ this ratio increases with the dimensionality of the system.
For large $U$ the ratio is similar for the two systems, since in both cases the band gap tends to $U$.

\section{Nickel oxide with the DER method\label{App:3}}
In Fig.\ \ref{figure7} we report the results for NiO obtained with the DER method.
In our calculations the Brillouin zone is sampled by a mesh of $6\times6\times6$ $\vect{k}$ points for
the paramagnetic phase and $4\times4\times4$ $\vect{k}$ points for the antiferromagnetic phase.
Both samplings include the $\Gamma$ point. Moreover, we used a smearing width of 27 meV.
In the antiferromagnetic phase the band gap is better reproduced than within the MEET.
This can be understood by inspecting the expression for removal/addition energies in the DER method.\cite{Sharma13} Using the power  functional these energies are obtained as:
\begin{align}
\epsilon^R_{i}=-\epsilon^A_{i}&=
\left.\frac{\partial E[\{n_k\},\{\phi_k\}]}{\partial n_{i}}\right|_{n_i=1/2}\nonumber\\
&= h_{ii}+\left[\sum_jV_{ijij}n_j-\alpha n^{\alpha-1}_{i}\sum_{j}V_{ijji} n^\alpha_{j}\right]_{n_i=1/2}.\label{Eqn:DER}
\end{align}
This expression, which is the same for both removal and addition energies, is similar to the expression for the removal energies $\delta_i^{R,(1)}$, which, with the power functional, reads
\begin{equation}
\delta^{R,(1)}_i =h_{ii}+\sum_jV_{ijij}n_j-n^{\alpha-1}_{i}\sum_{j}V_{ijji} n^\alpha_{j}.
\label{Eqn:delta_R}
\end{equation}
The difference between the two expressions resides in the prefactor $\alpha$ in the last term on the r.h.s. of Eq.\ \eqref{Eqn:DER}, which is equal to one in Eq.\ \eqref{Eqn:delta_R}, and the use of $n_i=1/2$ in  Eq.\ \eqref{Eqn:DER} instead of the value which minimizes the total energy $E[\{n_k\},\{\phi_k\}]$,\footnote{Note that in practice in the Elk code only the exchange-correlation contribution is evaluated at $n_i=1/2$.} as in Eq.\ \eqref{Eqn:delta_R}.  This tends to reduce the band gap width with respect to our method. Note that with $\alpha=1$ (HF), the two methods coincide.
In the paramagnetic phase {\color{black}(which we model with a non-magnetic phase)}, instead, we found that the DER method does not open any gap.
The occupation numbers mainly involved in the band gap region lie around 0.5 as in the MEET, however the corresponding energies accumulate in the band gap region, whereas our method displaces them and opens a gap. Note that our DER results are different from the results reported in Ref.\ \onlinecite{sharma_NiO}, where the DER method is shown to open a gap for a proper choice of the parameter $\alpha$. However, contrary to us, the authors of Ref.\ \onlinecite{sharma_NiO} use a shift of the $\vect{k}$-point grid. With the shift of the $\vect{k}$-point grid the number of inequivalent $\vect{k}$ points in the irreducible Brillouin zone is higher than in the case without shift and the calculations should converge faster. {\color{black}However, we found that this shift opens a gap even when used with HF, which, instead, should give a metal for paramagnetic NiO.\cite{Szpunar93} Moreover, a HF calculation without shift of the  $\mathbf{k}$-point grid on a slightly deformed crystal (deformation of the order of $10^{-9}$ relative to the lattice constant) yields a total energy lower than the case without deformation ($-1.58165 \cdot 10^3$ a.u. \textit{vs} $-1.58138\cdot 10^3$ a.u.), and a gapped spectral function very similar to the one obtained with the DER method when the shift of the grid is employed. 
Our analysis shows that these small perturbations lead to symmetry breaking through orbital ordering, and by consequence to the opening of a quasiparticle gap.
These findings suggest that the asymmetric shift of the $\mathbf{k}$-point grid used by Sharma \textit{et al}.\ \cite{sharma_NiO} plays an  important role in the appearance of a band gap in NiO and that there is no contradiction between our results and the ones published in Ref.\ \onlinecite{sharma_NiO}. It raises the question whether it is legitimate to break the orbital symmetry to model a paramagnetic system at zero temperature. This is clearly a very important issue, which deserves further investigations, but is beyond the scope of this work. 

In our calculations we did not use any shift of the $\mathbf{k}$-point grid and we checked that our results are converged with respect to the number of $\mathbf{k}$  points. We have also checked whether there is a starting point dependence. However starting from LDA or LDA+U yields the same conclusions.
Note that in LDA+U we do not break the spin symmetry, therefore  the $e_g$ bands are only shifted to higher energy and well separated by the $t_{2g}$ bands, but they remain degenerate. 
 
Of course the case of NiO is only one example, and this does not mean that the DER method does not open a gap in other paramagnetic transition metal oxides without breaking any symmetry. For example, in the case of MnO, the DER method opens a gap for certain values of $\alpha$ in the power functional.\cite{sharma_NiO} We checked that this is the case also without a shift of the $\mathbf{k}$-point grid.}

\bibliographystyle{apsrev}

\end{document}